\documentclass[twocolumn,preprintnumbers,amsmath,amssymb]{revtex4}
\usepackage{graphicx}

\usepackage{color}

\usepackage{dcolumn}
\usepackage{bm}

\newcommand{\comment}[1]{}

\newcommand{\yb}{$^{174}$Yb }
\newcommand{\ombar}{\bar{\omega}}

\begin{document}

\title{Rapid Cooling to Quantum Degeneracy in Dynamically Shaped Atom Traps}

\author{Richard Roy}
\author{Alaina Green}
\author{Ryan Bowler}
\author{Subhadeep Gupta}

\affiliation{Department of Physics, University of Washington, Seattle, Washington 98195, USA}

\date{\today}

\begin{abstract}
We report on a general method for the rapid production of quantum degenerate gases. Using $^{174}$Yb, we achieve an experimental cycle time as low as $(1.6\!-\!1.8)\,$s for the production of Bose-Einstein condensates (BECs) of $(0.5\!-\!1)\!\times\!10^5$ atoms. While laser cooling to $30\,\mu$K proceeds in a standard way, evaporative cooling is highly optimized by performing it in an optical trap that is dynamically shaped by utilizing the time-averaged potential of a single laser beam moving rapidly in one dimension. We also produce large ($>\!10^6$) atom number BECs and successfully model the evaporation dynamics over more than three orders of magnitude in phase space density. Our method provides a simple and general approach to solving the problem of long production times of quantum degenerate gases.
\end{abstract}
\maketitle

\section{\label{sec:intro} Introduction}

The production of quantum degenerate gases has revolutionized the field of atomic physics. Such gases are now routinely used as a means towards understanding complex many-body quantum phenomena from the realms of condensed matter and nuclear physics \cite{bloc08,bloc12}. As atom sources with precisely controlled properties, these gases can also significantly advance applications such as atom interferometry \cite{cron09} and quantum information processing \cite{jess04,schl11}. While the production and measurement methods of quantum gas experiments are well established, the measurement rate remains substantially limited by the lack of a general method for rapid sample production. Cycle times for such experiments are dominated by the production time, typically tens of seconds, while the actual experiment on the prepared sample lasts for about a second before destructive measurement. This separation of timescales is a severe impediment to the employment of quantum degenerate gases towards precision devices such as atomic clocks, inertial sensors and gravimeters \cite{dunn05,cron09,debs11}. Bridging these timescales can significantly contribute to all classes of quantum gas explorations and applications, as most measurements rely on the statistics of results from many experimental iterations.

The root of this timescale problem lies in the speed of collisional evaporative cooling. In a standard degenerate gas production sequence, the initial step of laser cooling produces temperatures in the few tens of $\mu$K, while light-induced processes keep the density below $10^{12}\,$cm$^{-3}$. The resulting phase space density (PSD) of $10^{-5}-10^{-4}$ is increased to quantum degeneracy by subsequent evaporative cooling in either magnetic or optical traps. Typical magnetic traps have large volumes and relatively low initial densities, yielding low collision rates and long evaporative cooling timescales of tens of seconds. The production of BECs in small-volume magnetic chip traps has provided one solution to the timescale problem \cite{hans01,fark10}. While this method has recently achieved cycle times of one second with $^{87}$Rb \cite{fark14}, it can only be applied to magnetic atoms.

Optical dipole traps (ODTs)\cite{barr01,gran02} provide the flexibility to cool all atoms, enabling applications with non-magnetic atoms and liberating the magnetic degree of freedom for interaction control during cooling \cite{gran02,webe03a}. Standard ODTs have small volumes and high collision rates, leading to smaller atom numbers but shorter evaporation timescales in the range of a few to $10\,$s. An overall BEC production time as low as $2\,$s has been accomplished in an ODT \cite{stel13SrProd} by combining one broad and one very narrow transition for laser cooling in $^{84}$Sr, to realize an extremely high initial PSD of $0.1$ before evaporative cooling \cite{footLCQD}. A BEC production time of $3.3\,$s was achieved in $^{87}$Rb by combining sub-Doppler laser cooling to a high initial PSD of $2 \times 10^{-3}$ with fast evaporative cooling in an ODT in which the volume was dynamically compressed by a moving lens \cite{kino05}. These methods however are not easily adaptable in a general way to other atomic species and experimental setups.



In this paper we present a general technique for rapid quantum degenerate gas production, where evaporative cooling is optimized by dynamically controlling the ODT shape with the time-averaged potential of a single rapidly-moving laser beam. This method straightforwardly allows for large initial and small final trap volumes, combining the advantages of earlier evaporative cooling strategies. It is applicable to all atoms and requires no hardware beyond a standard optical trapping setup. Applied to bosonic $^{174}$Yb with modest laser cooling to PSDs $<10^{-4}$, our experiment produces BECs containing $(0.5-1)\times 10^5$ atoms with an overall cycle time of $(1.6-1.8)\,$s. By suitably altering the time dependence of the trap parameters, we also produce large $^{174}$Yb BECs with $1.2\times10^6$ atoms. The observed evaporation dynamics are successfully captured over three orders of magnitude in PSD by our theoretical model.

\begin{figure}[t]
\includegraphics[width=1\columnwidth]{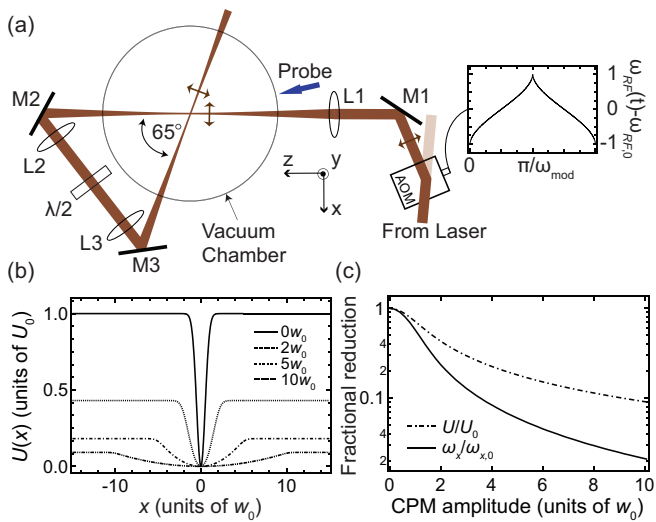}
\caption{\label{fig1} (a) Schematic of the optical trap setup. Gravity points along $y$. The horizontally oriented AOM is driven by an FM waveform (inset) resulting in parabolic intensity profile. Arrows on the laser beam indicate direction of CPM. (b) Evolution of time-averaged trap shape $U(x)=U(x,0,0)$ from Gaussian to parabolic with increasing CPM amplitude for fixed laser power in units of unmodulated beam waist $w_0$ and trap depth $U_0$. (c) Fractional reduction of trap depth and frequency in the painting direction $\omega_x$ from unmodulated values $U_0$ and $\omega_{x,0}$ for fixed power.}
\end{figure}

Forced evaporative cooling works by removing the high-energy tail of the Maxwell-Boltzmann distribution and allowing the remaining atomic sample to re-equilibrate by elastic collisions to a lower temperature. Keeping the removal point fixed at $\eta k_B T$ relative to the temperature $T$ allows the derivation of scaling laws for the evaporation dynamics \cite{kett96,ohar01}.
The rate of elastic collisions $\Gamma_{\rm el}=n_0\sigma \bar{v}$ and $\eta$ determine the per-particle evaporative loss rate as $\Gamma_{\rm ev}=\Gamma_{\rm el}(\eta - 4)e^{-\eta}$ for large $\eta$ \cite{ohar01,luiten96}, where $n_0$ is the peak particle density, $\sigma$ is the scattering cross section,  $\bar{v} = \sqrt{8k_B T/\pi m}$, and $m$ is the particle mass. We define the evaporation efficiency as $\gamma=-\frac{\ln(\rho_f/\rho_i)}{\ln(N_f/N_i)}$, where $\rho_{f(i)}$ and $N_{f(i)}$ are the final(initial) PSD and particle number, respectively. In the absence of additional loss processes, $\gamma$ can be made arbitrarily high by using a large $\eta$ and thus a long cooling timescale. The reality of other loss processes tempers this idea and introduces a new timescale which competes with that of evaporative cooling. This competition outlines an important experimental challenge and is captured by the ratio of ``good" to ``bad" collisions $R = \Gamma_{\rm el}/\Gamma_{\rm loss}$. The prescription for optimum efficiency however crucially depends on the nature of the dominant loss process.

For 1-body loss dominated systems characteristic of standard magnetic traps, $R$ is proportional to the elastic collision rate and scales as $N\ombar^3 T^{-1}$, where we have assumed a 3D harmonic trap with (geometric) mean trap frequency $\ombar$. Comparing against the scaling $\rho \propto N\ombar^3 T^{-3}$, we find that maintaining or increasing the collision rate at every step and achieving ``runaway" evaporation is an excellent prescription for efficient cooling \cite{kett96}. Importantly, this prescription simultaneously improves the speed of evaporative cooling and final particle number. In an ODT, while 3-body processes can be neglected in certain situations \cite{ohar01,hung08,deb14}, it is often the dominant loss mechanism. Then $R \propto N^{-1}\ombar^{-3} T^{2}$, and $\gamma$ and $R$ cannot be simultaneously optimized. Crucially, unlike in 1-body loss dominated systems, the inverted scaling of $R$ with density means maintaining a large $N$ leads to lowered $\ombar$, reduced collision rates and longer evaporation timescales. Numerical modeling of evaporative cooling in an ODT \cite{yanm11,olso13} can help optimize $\gamma$, but the challenges of large number and speed remain.

Our solution to these challenges involves the implementation of the time-averaged optical potential of a laser beam moving rapidly, or ``painting", in one dimension. The ability to dynamically control the center position modulation (CPM) amplitude of the beam in addition to its total power, results in independent, arbitrary control over both the trap depth ($U=\eta k_B T$) and frequency as a function of time, a key advantage over methods with a fixed power-law relationship between $U$ and $\bar{\omega}$ \cite{ohar01,olso13}.

The rest of the paper is organized as follows. Section \ref{sec:expset} consists of the experimental setup and our model for the CPM trap shape. In Section \ref{sec:evapmodel} we present our numerical model for forced evaporative cooling in this trap along with a procedure for optimizing evaporation efficiency, and apply our model to an experimentally optimized evaporation trajectory. In Section \ref{sec:rapidBECs} we describe rapid BEC production using our method, while in Section \ref{sec:largeBECs} we focus on the production of large BECs. Finally, Section \ref{sec:summary} provides a summary of this technique and an outlook for applications.

\section{\label{sec:expset} Experimental setup}

We perform our experiment in the apparatus described in \cite{hans13} using \yb bosons. Our laser cooling procedure can produce $10^8$ atoms in $5\,$s at $30\,\mu$K in a compressed magneto-optical trap (MOT) operating on the $^1S_0\! \rightarrow ^{3}\!\!P_1$ transition. In the rest of this section, we describe the setup of our ODT and characterization of the parabolic CPM trap.

\subsection{\label{subsec:trapdetails} Optical trap details}

The ODT (Fig.\,\ref{fig1}) is generated by sending the output of a fiber laser at 1064 nm (IPG YLR-100-LP) through an acousto-optic modulator (AOM, 80 MHz, Intraaction ATM-804DA6B) and focusing the diffracted beam (1$^{\text{st}}$ order) to a Gaussian waist of 35$\,\mu$m at the atoms. This light is then refocused with orthogonal polarization back onto the atoms with a waist of 30$\,\mu$m, at an angle of 65$^\circ$ with respect to the first pass.

To implement the trap center position modulation, we modulate the center frequency of the voltage controlled oscillator (VCO, Minicircuits ZOS-100) supplying RF to the AOM at 10 kHz with the waveform shown in Fig.\,\ref{fig1}(a), which results in a nearly perfect parabolic trap shape (see Appendix). The largest CPM amplitude used is 260 $\mu$m (520 $\mu$m peak-to-peak), and corresponds to shifting the center frequency of the AOM by 7 MHz (14 MHz peak-to-peak) on top of 80 MHz. The orientation of the AOM is such that the CPM occurs in the horizontal plane. We control the overall ODT power with the RF drive strength to the AOM. Using a power of $70\,$W and CPM amplitude of $260\,\mu$m at the atoms, we capture up to $5 \times 10^7$ atoms from the compressed MOT into the ODT.

\subsection{\label{subsec:trapchar} CPM trap characterization}

\begin{figure}
\centering
\includegraphics[width=1\columnwidth]{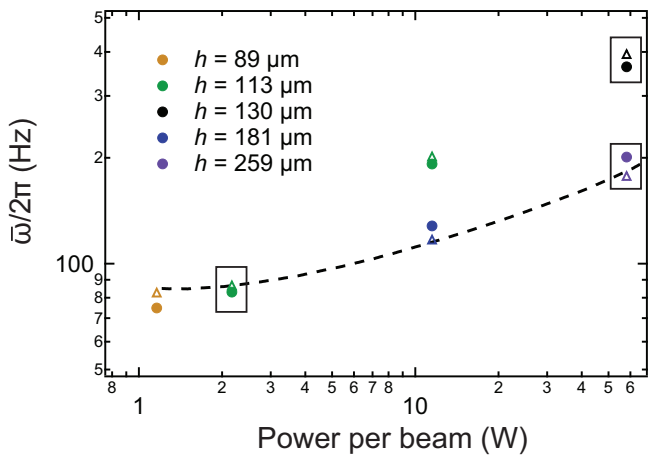}
\caption{\label{fig2} Comparison of CPM trap model (open triangles) to measurements (solid circles) of the trap frequency. Different CPM amplitudes are indicated with different colors. Boxes indicate trap configurations used for measurements in Fig. \ref{fig3}. The black dashed line is a parametric plot of $(\bar{\omega} (P(t),h(t))/2\pi, P(t))$ corresponding to the evaporation trajectory used in Fig. \ref{fig5}.}
\end{figure}

Crucial to the implementation of our forced evaporative cooling model (see Section \ref{sec:evapmodel}) is an accurate model for the optical trap shape as a function of laser power $P$ and CPM amplitude $h$ \cite{foothdef}. Without CPM and neglecting gravity, one can apply the scalings $U(P) \propto P$ and $\bar{\omega}(P) \propto P^{1/2}$ in the absence of beam imperfections (e.g. thermal lensing, astigmatism, etc.). However, as depicted in Fig. \ref{fig1}, CPM allows for a large range of depths and frequencies at each power. We calculate the trap frequencies and trap depth for a single ODT beam with CPM in the $x$ direction and traveling in the $z$ direction as
\begin{align} \label{eqn:cpmfreqsanddepth}
\omega_{x}^2 = \frac{8\alpha P}{\pi m w_0^4}f_\omega(h/w_0),& \hspace{.2in} \omega_z^2 = \frac{4\alpha P}{\pi m w_0^2 z_R^2}f_U(h/w_0), \nonumber \\
\omega_y^2 = \frac{8\alpha P}{\pi m w_0^4}f_U(h/w_0),& \hspace{.2in} U = \frac{2\alpha P}{\pi w_0^2}f_U(h/w_0),
\end{align}
where $\alpha$ is the atomic polarizability at the ODT wavelength $\lambda$ and $z_R=\pi w_0^2/\lambda$ is the Rayleigh range. $f_{\omega}(h/w_0)$ and $f_{U}(h/w_0)$ are the fractional reduction factors shown in Fig. \ref{fig1}(c), conveniently written as a function of the CPM amplitude in units of beam waist.

To arrive at a reliable model for our specific time-averaged potential, we measure the physical amplitude $h$ (in $\mu$m) of the center position modulation at the ODT focus for a given applied voltage to the VCO. Furthermore, we account for the effects of ellipticity and thermal lensing by allowing power dependent waists $w_{x} = w_{x}(P)$ and $w_{y} = w_{y}(P)$, changing the arguments of the fractional reduction functions to $f_U(h/w_{x}(P))$ and $f_\omega(h/w_{x}(P))$, and appropriately altering the equations in (\ref{eqn:cpmfreqsanddepth}).

\begin{figure}
\centering
\includegraphics[width=1\columnwidth]{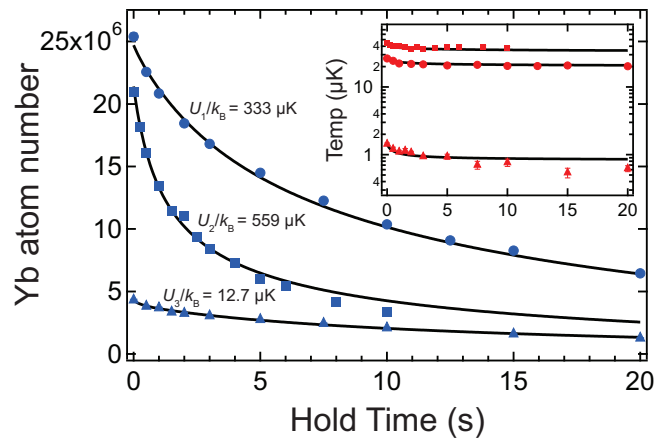}
\caption{\label{fig3} Measurements of Yb number and temperature evolution for trap depth model calibration. The 3 different trap configurations are as follows: (1, solid circles) $P$ = 58 W, $h$ = 259 $\mu$m; (2, solid squares) $P$ = 58 W, $h$ = 130 $\mu$m; (3, solid triangles) $P$ = 2.2 W, $h$ = 113 $\mu$m. The lines are fits using the evaporation model discussed in Section \ref{sec:evapmodel}. The extracted trap depth for each curve is indicated on the plot above. Our time-averaged trap model predicts $U_1/k_B = 300$ $\mu$K, $U_2/k_B = 576$ $\mu$K, and $U_3/k_B = 13.3$ $\mu$K.}
\end{figure}


Finally, we include the second pass of the crossed dipole trap by using a magnification of 5/6 given by the lenses used to reimage the beam back onto the atoms at 65$^\circ$. We write $U_\text{total}(x,y,z) = U_1(x,y,z) + U_2(x',y',z') + mgy$, where the subscripts 1 and 2 refer to the first and second passes of the beam, and the coordinates $(x',y',z')$ are related to $(x,y,z)$ by a 65$^\circ$ rotation in the $x-z$ plane. We then compute the roots of $\partial U_\text{total}(0,y,0)/\partial y \vert_{y=y^*} = 0$ for a densely spaced grid of $P$ and $h$ values, and use these roots to compute $U(P,h)$ and $\bar{\omega}(P,h)$ referenced to the equilibrium position $y^*$.

The comparison of our trap model to trap frequency measurements made using $^{174}$Yb is shown in Fig. \ref{fig2}. As discussed above, all trap parameters are independently determined by observations of the beam itself, so there are no fit parameters used here. We find good agreement between the theoretical and experimental values.

To compare our model for the trap depth with the experiment, we perform measurements of Yb number and temperature evolution for three different fixed settings of ODT power and CPM amplitude (indicated in Fig. \ref{fig2} with boxes). We then fit the observed dynamics with our evaporation model (see Section \ref{sec:evapmodel}) for fixed power and CPM amplitude, using the trap depth as a fitting parameter. The results are shown in Fig. \ref{fig3}. The predicted trap depth values from our model are given in the caption to Fig. \ref{fig3}, and are in good agreement with those extracted from number and temperature dynamics.

\section{\label{sec:evapmodel} Evaporation model}

We model the number and temperature evolution in the high $\eta$ limit, where the equation of state is well approximated by $E=3Nk_B T$. The dynamical equations are then \cite{luiten96,ohar01,yanm11}
\begin{gather}
\dot{N} = -(\Gamma_\text{ev} + \Gamma_\text{3b} + \Gamma_\text{bg})N \label{ndot} \\
\dot{T} = -\left(\frac{\Gamma_\text{ev}}{3}\left(\eta +\alpha - 3 \right) - \frac{\Gamma_\text{3b}}{3} - \frac{\dot{\bar{\omega}}}{\bar{\omega}} \right) T + \frac{\Gamma_\text{sc} E_\text{r}}{3k_B}, \label{tdot}
\end{gather}
where $\Gamma_\text{3b} = K_3 N^{-1} \int n^3 d^3\vec{r}$ is the per-particle loss rate for 3-body inelastic loss, $K_3$ is a temperature-independent 3-body inelastic loss rate coefficient, $\alpha = (\eta-5)/(\eta-4)$, and $\Gamma_\text{sc}$ and $E_\text{r}$ are the spontaneous scattering rate and recoil energy for $^{174}$Yb in our 1064 nm ODT. For the ODT intensities used here, heating from spontaneous scattering is small. The background lifetime $\Gamma_\text{bg}^{-1}$ is independently measured to be 35$\,$s.


\subsection{Optimization of evaporation trajectory}

By neglecting the effects of 3-body inelastic loss, background gas collisions, and spontaneous scattering, one can construct analytical solutions to equations (\ref{ndot})-(\ref{tdot}) based on scaling laws by specifying a trajectory $\eta(t) = 10$ for all times $t$ and assuming a fixed relationship $\bar{\omega} \propto P^{1/2}$ \cite{ohar01}. Without these simplifying assumptions, however, it is essential to turn to numerical techniques, especially with an additional dynamically controllable parameter such as CPM amplitude. Therefore, to inform our experimental choice of ramp profiles $P(t)$ and $h(t)$, we run an optimization algorithm in Mathematica$^\text{TM}$ using numerical solutions to equations (\ref{ndot})-(\ref{tdot}). To speed up the numerical integration of $\dot{N}$ and $\dot{T}$, we compute the trap parameters $U(P,h)$ and $\bar{\omega}(P,h)$ for a dense grid of power and CPM amplitude values. We then convert these tables into interpolating functions, making the determination of $U(t)$ and $\bar{\omega}(t)$ throughout the evaporation ramp extremely fast.

For the functional form of the power ramp $P(t)$ we try single exponential profiles of the form $P(t) = P_0e^{-t/\tau_P}$ and bi-exponential profiles of the form $P(t) = P_0\left(\alpha e^{-t/\tau_{P1}} + (1-\alpha) e^{-t/\tau_{P2}}\right)$, guided by the fact that evaporation occurs on an exponential timescale. Furthermore, a bi-exponential could potentially handle the presence of two dominant timescales (e.g. evaporation and 3-body loss). For the CPM amplitude ramp $h(t)$, we try out many different functional forms, including offset exponential $h(t) = h_1e^{-t/\tau_h} + (h_0-h_1)$, linear $h(t) = \text{max}(h_0-\beta t,0)$,  and concave functions $h(t) = \text{max}(h_0 - h_1(1-e^{t/\tau_h}),0)$ and $h(t) = \text{max}(h_0 - (t/\tau_h)^2,0)$. For the above functions, the parameters $P_0$ and $h_0$ are fixed as they correspond to the trap loading conditions, while the remaining parameters are varied as part of the optimization algorithm.

Our optimization algorithm utilizes the gradient ascent method where $\gamma = -\frac{\ln (\rho(t_f)/\rho(0))}{\ln (N(t_f)/N(0))}$ is the quantity to be maximized, and $t_f$ satisfies $\rho(t_f) = 1$. For the power ramp profile, we find that the optimization procedure pushes the bi-exponential profile towards a single exponential (i.e. $\tau_{P1} \rightarrow \tau_{P2}$). CPM ramp optimization suggests that the system is very robust to the form of $h(t)$. In fact, if we use $P(t) = P_0 e^{-t/\tau_P}$ with $\tau_P = 1\,$s, and run the optimization algorithm for each of the proposed functions $h(t)$ above, the optimized values $\gamma_\text{opt}$ are all within 1\% of each other. For these reasons we choose to adopt the single exponential power ramp and linear CPM ramp, as these are the simplest options.

\begin{figure}[b]
\centering
\includegraphics[width=1\columnwidth]{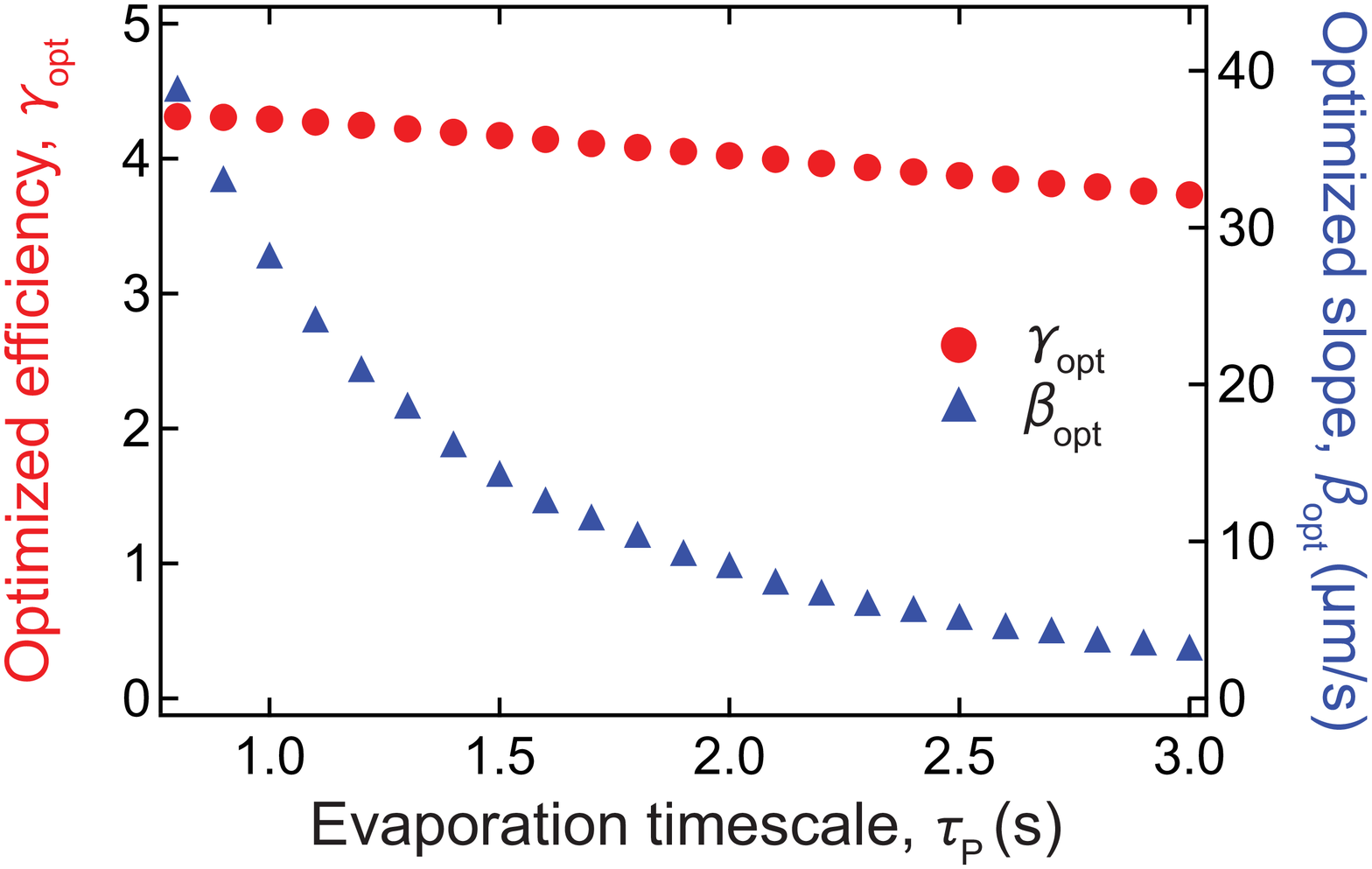}
\caption{\label{fig4} CPM amplitude trajectory optimization for various power ramp timescales $\tau_P$. For each value $\tau_P$ we run an optimization of the CPM reduction slope $\beta$ (solid blue triangles) to maximize the evaporation efficiency $\gamma$ (solid red circles) at $\rho = 1$.} 
\end{figure}

We find the system to be extremely robust to the timescale of evaporation, as indicated in Fig. \ref{fig4}. For these simulations, we fix the timescale $\tau_P$ and optimize the slope of the CPM ramp $\beta$. We restrict $\tau_P \gtrsim 0.8$, as we believe that our model cannot accurately capture the evaporation dynamics on timescales faster than this due to complications involving thermal lensing and decoupling of horizontal and vertical temperatures. As seen in Fig. \ref{fig4}, the optimized evaporation efficiency $\gamma_\text{opt}$ varies little over the range $0.8 \leq \tau_P \leq 3$. In fact, the slight slope of $\gamma_\text{opt}$ versus $\tau_P$ is caused by the introduction of a new timescale, the background lifetime $\Gamma_\text{bg}^{-1}$. As described in Section \ref{sec:comparemodandexp}, we observe the same behavior in the experiment as the maximum number of atoms in the Yb BEC is quite resilient to the evaporation timescale.

\subsection{\label{sec:comparemodandexp} Comparison of model and experiment}

\begin{figure}[t]
\centering
\includegraphics[width=1\columnwidth]{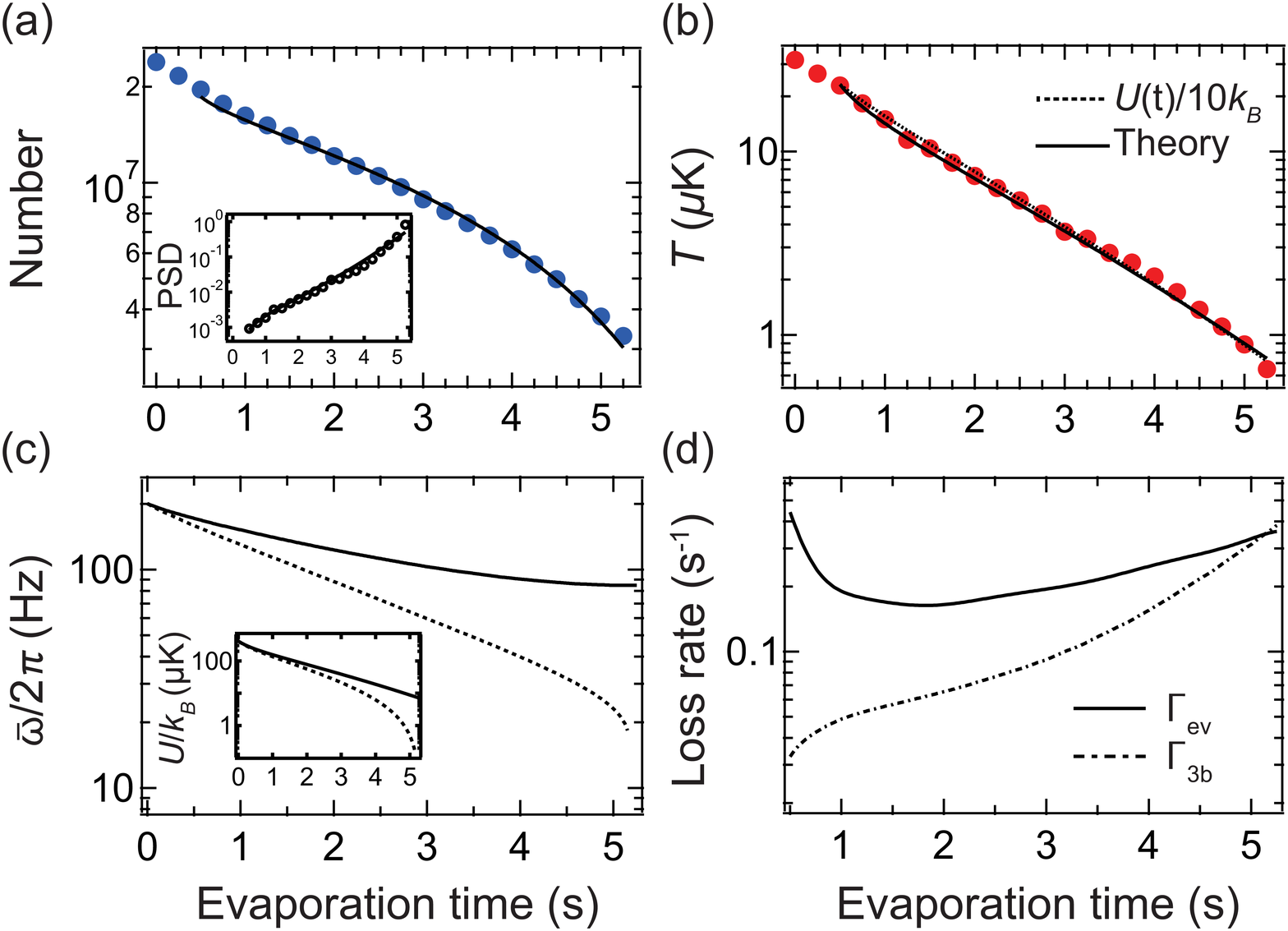}
\caption{\label{fig5} Example of optimized evaporation trajectory. (a,b) Measured number (blue circles), temperature (red circles), and phase space density (black circles, inset) evolution during forced evaporative cooling show excellent agreement with our theoretical model (solid black line, see text). In (b) we also plot $U(t)/10k_B$ (dashed line). From the fit in (b) we find $\eta_\text{avg} = 10.5$. (c) Trajectories of trap frequency and depth (inset) for measurements in (a,b). Dashed lines correspond to the same trajectories without reduction of CPM amplitude. (d) Evolution of per-particle loss rates for evaporation (solid line) and 3-body inelastic loss (dashed line) during forced evaporation.}
\end{figure}

We experimentally investigate the evaporation efficiency by maximizing final BEC number. In agreement with our simulations we find that over timescales where 1-body loss is negligible, the evaporation efficiency is robust and an exponential reduction of power and linear reduction of CPM amplitude yield the largest $\gamma$. A typical optimized evaporation trajectory is shown in Fig.\,\ref{fig5}. Our theoretical model successfully captures the dynamics over 3 orders of magnitude in PSD. For these measurements, we wait $500\,$ms after ODT loading before beginning forced evaporation to allow atoms in the wings of the trap to escape.

For the model curves in Fig.\,\ref{fig5}, the only free parameters are the initial number and temperature, and $K_3$. We assume $s$-wave scattering only as the $d$-wave threshold for \yb is 75 $\mu$K, and use $\sigma = 8\pi a^2$, where $a = 5.6$ nm \cite{kita08}. From a least squares fit to the data, we extract $K_3 = (1.08 \pm 0.03) \times 10^{-28}$ cm$^6$s$^{-1}$. From the behavior of $\Gamma_\text{ev}$ and $\Gamma_\text{3b}$ in Fig.\,\ref{fig5}(d), we see that the dynamically shaped ODT maintains dominance of evaporative over 3-body loss. Furthermore, the elastic collision rate $\Gamma_\text{el}$ falls less than a factor of 2 from 1.7 kHz to 1.1 kHz over the course of the evaporation sequence. The evaporation efficiency $\gamma$ for this ramp is 3.8, close to the highest value found using the optimization algorithm discussed above with our initial trap conditions and ramp profiles. We note that although runaway evaporation where $d\Gamma_\text{el}/dt >0$ is easily achieved with the dynamically shaped ODT, we do not find this to be the optimal evaporation strategy due to enhanced 3-body loss.

\section{\label{sec:rapidBECs} Rapid BEC production}

\begin{figure}[t]
\centering
\includegraphics[width=1\columnwidth]{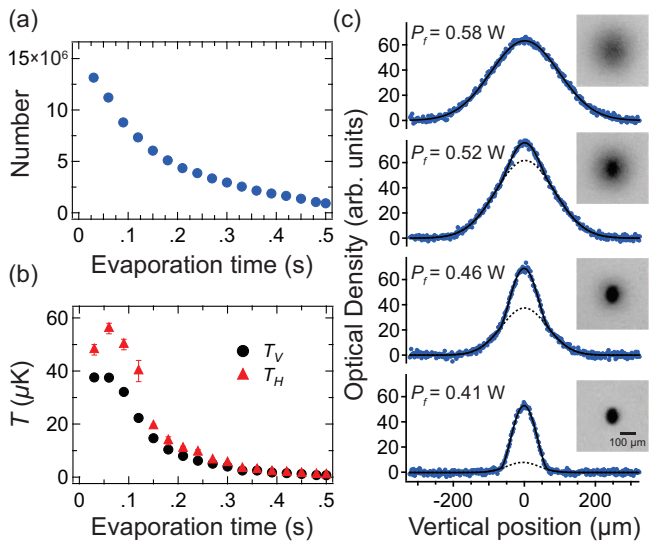}
\caption{\label{fig6} Rapid production of $^{174}$Yb BECs with a total cycle time of $1.8\,$s. (a,b) Evolution of number and horizontal ($T_H$) and vertical ($T_V$) temperatures during first 2 phases of rapid evaporation (see text). (c) Absorption images (insets) and horizontally integrated OD profiles after the 3rd phase of evaporation and 25 ms time-of-flight (ToF) to a variable final laser power $P_f$. Solid lines are fits to a bimodal density distribution (see text), with dashed lines indicating fits to the thermal component. For $P_f = 0.41$ W, we find a nearly pure condensate of $1 \times 10^5$ atoms.}
\end{figure}

In addition to providing a platform for highly efficient evaporation of large atom number clouds, dynamical trap shaping can be applied to the rapid production of BECs. For this purpose we shorten the MOT loading and compression time to a total of $0.8\,$s, and begin forced evaporative cooling immediately following loading of the ODT with an initial PSD of $<10^{-4}$. The experimentally optimized fast BEC ramp measurements are shown in Fig.\,\ref{fig6}.

For fast BEC production, we find the use of three distinct evaporation stages to be optimal. In the first stage, we exponentially reduce the power by a factor of 20 in $200\,$ms while linearly reducing the CPM amplitude to zero in $150\,$ms. Second, we exponentially reduce the laser power by a factor of 6 in $300\,$ms. In a third, relatively slow evaporation stage, we linearly decrease the power another 30\% in $350\,$ms, and then hold at constant depth for $150\,$ms. In Fig.\,\ref{fig6}(b) we see that the horizontal and vertical temperatures initially decouple due to the rapid decrease of CPM amplitude in the first $150\,$ms. This can be understood by considering the adiabatic temperature evolution terms, since $\dot{\omega}_{x,z}/\omega_{x,z} \gg \dot{\omega}_y/\omega_y$ during this time. Since the assumption of thermal equilibrium is violated at these short timescales, we cannot apply our model in this regime.

Fig.\,\ref{fig6}(c) shows absorption images and horizontally integrated optical density (OD) profiles after the third evaporation stage for a few final laser powers near the condensation transition. We fit the density profiles to a bimodal distribution consisting of Gaussian thermal and Thomas-Fermi BEC profiles. We detect nearly pure $^{174}$Yb condensates of $1 \times 10^5$ atoms for $P_f = 0.41\,$W, with a total cycle time of $1.8\,$s. By shortening the MOT loading and compression time to $0.6\,$s, we produce BECs of $0.5 \times 10^5$ atoms with a total cycle time of $1.6\,$s.

\section{\label{sec:largeBECs} Application to large BEC production}

We now turn our attention to the production of large atom number condensates. As shown in Fig.\,\ref{fig5}(d), the 3-body loss rate grows noticeably near the point $\rho = 1$ due to a large increase in the density. Therefore, in order to produce the largest condensates we evaporate with the same functional ramps as in Fig.\,\ref{fig5} until $\rho \approx 1$, and subsequently continue forced evaporation by fixing the power and increasing the CPM amplitude.

\begin{figure}[t]
\centering
\includegraphics[width=1\columnwidth]{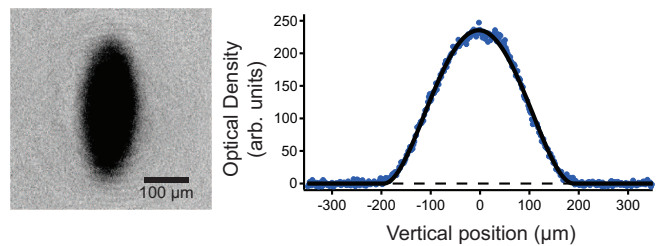}
\caption{\label{fig7} Production of large, pure $^{174}$Yb BECs using dynamical trap shaping to combat 3-body inelastic loss. Absorption image (left) and horizontally integrated OD profile (right) after 50 ms ToF, with fit (solid line) to a Thomas-Fermi BEC density distribution, yielding a total of $1.2 \times 10^6$ condensed atoms. The thermal fraction is consistent with zero.}
\end{figure}

Fig.\,\ref{fig7} shows an example absorption image and integrated OD profile of a pure BEC produced from such an evaporation ramp. For this particular measurement we finish the initial evaporation stage at $\rho \approx 1$ with $h = 130\,\mu$m and $P = 1.0$ W, and subsequently increase the CPM amplitude to $h = 180\,\mu$m. The resulting trap frequencies are $\left(\omega_x,\omega_y,\omega_z\right) = 2\pi \times \left(17,110,10\right)$ Hz. Following this method we can reliably create pure $^{174}$Yb condensates of $1.2 \times 10^6$ atoms, a factor of 4 improvement over the largest reported Yb BEC number \cite{hans13}, with a total cycle time of $15\,$s.

\section{\label{sec:summary} Summary and outlook}

While an additional CPM degree of freedom from a second orthogonal AOM can allow more control over atom cloud compression \cite{bien12} or final BEC shape \cite{hend09,ryuc13}, it is unlikely that it will provide significant improvements to the speed and efficiency of evaporative cooling that we have demonstrated here. In our current implementation, we choose the vertical waist and large initial CPM amplitude to load about 50\% of the compressed MOT. Once a large atom number has been loaded, we control all the relevant parameters for efficient evaporative cooling by dynamically shaping the trap (Fig.\,\ref{fig5}) to provide an appropriate $\ombar$ to maximize $\gamma$ and simultaneously provide large $\omega_y$ for tight confinement against gravity. 

The CPM amplitude provides a simple way to control the trap aspect ratio over a large range. Furthermore, a large CPM amplitude with $\omega_y \gg \omega_{x,z}$ can be used to realize 2D confinement, and square-wave modulation can be used to generate multi-well traps.

Our fast and efficient quantum gas production methods can be applied fruitfully to other atomic species, fermions and mixtures, and can significantly impact various applications, such as atom interferometry \cite{cron09,debs11,jami14}. The demonstrated fast BEC production time in this work is at least one order of magnitude shorter than that in typical quantum degenerate gas experiments. The leading limitation to our cycle time is the MOT loading rate, stemming from the relatively narrow linewidth of the $^1S_0\!\rightarrow\!\! {^3P_1}$ transition. Combining broad- and narrow-line laser cooling either spatially \cite{leej15} or temporally \cite{maru03} can shorten the cycle time further, and is also applicable to other alkaline-earth \cite{kraf09,stel13SrProd,mart09} as well as lanthanide \cite{berg08,lubu11,fris12} atoms. Our scheme should also improve the cycle time of alkali atom experiments which feature the advantage of sub-Doppler cooling with relatively broad transitions as well as the combination of broad and narrow transitions \cite{duar11,mcka11}. 

The broad applicability of our method to all laser-cooled atomic species gathers additional appeal when one considers that most optical trapping experiments already involve one AOM to control the power during evaporative cooling. The short cycle time can also allow for further technical simplifications including reduced vacuum requirements and the use of lower ODT powers at wavelengths closer to atomic resonance. In alkali atoms such as Rb and Cs, where laser cooling can produce samples at a few $\mu$K with PSD $> 10^{-3}$ \cite{kerm00,kino05}, an order of magnitude lower initial trap depth than used here is adequate. Estimating for the commonly used $^{87}$Rb atom, where $\sigma$ and $K_3$ \cite{burt97} are similar to $^{174}$Yb, it should be possible to have $1\,$s BEC production times with only $0.5\,$W of ODT power at $100\,$nm detuning.

\begin{acknowledgments}
We thank Chao Yi Chow for various technical contributions to the experiment. We gratefully acknowledge financial support from NSF Grant No. PHY-1306647, AFOSR Grant No. FA 9550-15-1-0220, and ARO MURI Grant No. W911NF-12-1-0476.
\end{acknowledgments}

\renewcommand{\theequation}{A\arabic{equation}}
\setcounter{equation}{0}  
\section*{APPENDIX: FM WAVEFORM DERIVATION}

\begin{figure}
\centering
\includegraphics[width=1\columnwidth]{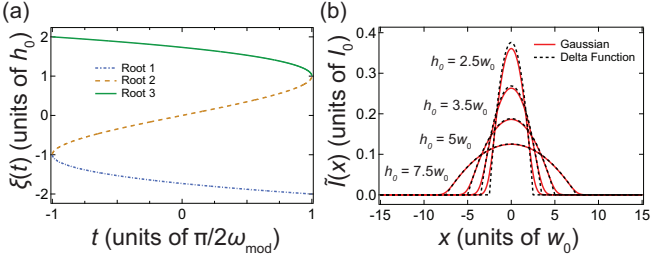}
\caption{\label{fig8} FM waveform for parabolic time-averaged potential. (a) The 3 roots of the implicit equation $\xi^3/3h_0^2-\xi+t=0$ plotted in the domain $-\pi/2\omega_\text{mod} \leq t \leq \pi/2\omega_\text{mod}$, corresponding to a half period. The root of interest for CPM is centered on the $\xi$ axis (Root 2). (b) Time-averaged CPM potentials using root 2 from (a) as the FM waveform for delta function (dashed black line) and Gaussian (solid red line) initial beams.}
\end{figure}

In order to derive the FM waveform necessary to create the parabolic time-averaged potential shown in Fig. 1(b), we work with 1D transverse profiles in the painting dimension, $x$. Consider the unmodulated beam shape to be a delta function $I(x) = P_0 \delta(x)$. In this case, for a CPM amplitude of $h_0$ we want the time-averaged intensity distribution to be of the form
\begin{align}
\tilde{I}(x) = \frac{3P_0}{4h_0^3}(h_0^2-x^2)\Theta(h_0-|x|), \label{eqn:timeaveintensity}
\end{align}
where $\Theta$ is the Heaviside function. To realize this potential we want to find the function $\xi(t)$ such that the time average of $I(x-\xi(t))$ from $t_1$ to $t_2$ equals the expression in equation (\ref{eqn:timeaveintensity}), where $\xi(t_1)=-h_0$ and $\xi(t_2) = h_0$.

We reason that the rastered delta function must spend an amount of time at each point $\xi'$ that obeys
\begin{align}
\frac{dt\rvert_{\xi=\xi'}}{dt\rvert_{\xi=0}} = \frac{\tilde{I}(\xi')}{\tilde{I}(0)} = \left(1-\left(\frac{\xi'}{h_0}\right)^2\right).
\end{align}
Writing $dt$ as $d\xi/\dot{\xi}$ we find
\begin{align}
\frac{dt\rvert_{\xi=\xi'}}{dt\rvert_{\xi=0}} = \frac{\dot{\xi}\rvert_{\xi=0}}{\dot{\xi}\rvert_{\xi=\xi'}} = \left(1-\left(\frac{\xi'}{h_0}\right)^2\right).
\end{align}
Treating $\dot{\xi}\rvert_{\xi=0} \equiv v_0$ as a constant determined by $\omega_\text{mod}$ and $h_0$, we arrive at a differential equation for $\xi(t)$
\begin{align}
\frac{d\xi}{dt} = \frac{v_0}{1-\left(\frac{\xi}{h_0}\right)^2}, \hspace{.2in} |\xi| \leq h_0. \label{forcepsode}
\end{align}
Solving equation (\ref{forcepsode}) gives the implicit equation $\xi(t)^3/3h_0^2 - \xi(t) + v_0t = 0$.  The 3 roots are plotted in Fig. \ref{fig8}(a). Clearly the solution that is centered on the y axis will be the desired root. From the constraint $|\xi| \leq h_0$, we find $|t| \leq 2h_0/3v_0$. Anticipating that we will construct the periodic waveform shown in the inset to Fig. \ref{fig1}(a), we define the modulation period $2\pi/\omega_\text{mod} = 8h_0/3v_0$.  Next we compute the time average with $t_1 = -2h_0/3v_0$ and $t_2 = 2h_0/3v_0$,
\begin{align}
\tilde{I}(x) &= \frac{3v_0P_0}{4h_0} \int\limits_{-2h_0/3v_0}^{2h_0/3v_0} \delta(x-\xi(t)) dt \nonumber \\
&= \frac{\omega_\text{mod}P_0}{\pi} \int\limits_{-\pi/2\omega_\text{mod}}^{\pi/2\omega_\text{mod}} \frac{\delta(t-t_0)}{|\dot{\xi}(t_0)|} dt,
\end{align}
where $t_0$ satisfies $x -\xi(t_0) = 0$. From equation (\ref{forcepsode}) we have $\dot{\xi}(t_0)=v_0/(1-(\xi(t_0)/h_0)^2)$. Lastly, the integral over the delta function evaluates to zero unless $|t_0| \leq 2h_0/3v_0$, which is equivalent to $|\xi(t_0)| \leq h_0$. Therefore,
\begin{align}
\tilde{I}(x) &= \frac{3P_0}{4h_0} \left(1-(\xi(t_0)/h_0)^2\right) \Theta(h_0-|\xi(t_0)|) \nonumber \\
&= \frac{3P_0}{4h_0^3} \left(h_0^2-x^2\right) \Theta(h_0-|x|).
\end{align}

Fig. \ref{fig8}(b) shows the time-averaged potentials for delta function $I(x) = P_0 \delta(x)$ and Gaussian $I(x) = I_0 \exp(-2x^2/w_0^2)$ initial beam shapes, where $I_0 = \sqrt{2/\pi}P_0/w_0$. As seen in the figure, the cases of the delta function and Gaussian differ very little, and become indiscernible when $h_0$ is much greater than the Gaussian waist. In order to perform frequency modulation with the waveform $\xi(t)$ corresponding to the second root in Fig. \ref{fig8}(a), we utilize the arbitrary waveform functionality of a Stanford Research Systems DS345 function generator, passing it an array of values very closely approximating the periodic continuation of $\xi(t)$ (see inset to Fig. \ref{fig1}).



\begin{thebibliography}{42}
\bibitem[{foo(2015{\natexlab{a}})}]{footLCQD}
\bibinfo{journal}{We note that laser cooling to quantum degeneracy has also
  been demonstrated in $^{84}$Sr \cite{stel13}, but with an experimental cycle
  time of $10\,$s for pure BEC production.}
\bibitem[{foo(2015{\natexlab{b}})}]{foothdef}
\bibinfo{journal}{The quoted values of $h$ correspond to that for the first
  focus of the ODT. We scale $h$ by the ratio of the beam waists for the second
  beam focus.}
\end{thebibliography}
\end{document}